\documentclass[journal]{IEEEtran}

\usepackage{cite}

\ifCLASSINFOpdf
   \usepackage[pdftex]{graphicx}
		\usepackage{epstopdf}
\else
   \usepackage[dvips]{graphicx}
\fi
\usepackage{amsmath}
\ifCLASSOPTIONcompsoc
  \usepackage[caption=false,font=normalsize,labelfont=sf,textfont=sf]{subfig}
\else
  \usepackage[caption=false,font=footnotesize]{subfig}
\fi

\usepackage{siunitx}
\usepackage{tikz}
\usetikzlibrary{patterns}

\begin{document}
%
\title{Drain Current Model of One-Dimensional Ballistic Reconfigurable Transistors}
%
%
%

\author{Igor~Bejenari 
				\thanks{ This work was supported in part by DFG project CL384/2 and  DFG project SCHR695/6.}%
\thanks{I. Bejenari is with the Chair for Electron Devices and Integrated Circuits, Department of Electrical and Computer Engineering, 
 Technische Universit\"at Dresden, 01062, Germany.}%
\thanks{I. Bejenari is  also with Institute of Electronic Engineering and Nanotechnologies,  Academy of Sciences of Moldova, MD 2028 Chisinau, Moldova (e-mail:igor.bejenari@fulbrightmail.org).}
				
}%

%
%

\markboth{}%
{Bejenari \MakeLowercase{\textit{et al.}}: Analytical Model of One-Dimensional Ballistic Schottky-Barrier Transistors}
\maketitle
\begin{abstract}
A simple model based on the WKB approximation for one-dimensional ballistic multi--gate reconfigurable field--effect transistors (RFETs) with Schottky-Barrier  contacts has been developed  for the drain current taking into account electron and hole band-to-band tunneling. By using a proper approximation of both the Fermi-Dirac distribution function and transmission probability, an analytical solution for the Landauer integral can be obtained. A comparative analysis of the two-gate and triple-gate RFETs is performed based on the numerical integration of the current integral. 
\end{abstract}


%
\IEEEpeerreviewmaketitle

\section{Introduction}
%
%
%
%
\IEEEPARstart{S}{ome} of the recent requirements for CMOS technology listed in the International
Roadmap for Devices and Systems (IRDS) \cite{IRDS} include  high--mobility channel materials, 
gate--all--around (nanowire) structures, scaling down  supply voltages lower than 0.6 V, controlling source/drain series resistance within tolerable limits, providing lower Schottky--barrier (SB) height, and fabrication of advanced nonplanar multi--gate and nanowire MOSFETs. 
Along with FETs based on semiconductor nanowires, carbon-nanotube FETs (CNTFETs) satisfy these requirements \cite{Guo_IEEE2004, Franklin_2012,Peng2017}. 
Downscaling the transistor dimensions goes along with a transformation of ohmic contacts into Schottky contacts~\cite{Larson_IEEE2006,Leonard2011}.
Due to a possible low channel resistance (or even ballistic conduction), the metal-semiconductor contact
resistance can significantly affect or even dominate the performance of SB transistors~\cite{Chen_IEEE2006,Heinze2003,Chen2005}.
In contrast to conventional FETs, multi-gate  reconfigurable field--effect transistors (RFET) can be configured between an n-- and p--type by applying an electrical signal, which selectively controls  charge carrier injections at each Schottky contact, explicitly avoiding the material doping~\cite{Weber2017,Mikolajick2017}.
RFETs have the potential to enable adaptive and reconfigurable electronics, which can lead to the initiation of
radically new circuit paradigms and computing schemes based on the reprogrammable logic with the reduced number of required devices.
Along with the electron tunneling through SB, the band-to-band tunneling (BTBT)  of electrons  has significant effect on RFET characteristics.
This  leads to an increase of current and decrease of a subthreshold swing, which can be less than its limit value of 60 mV/dec  typical for MOSFETs at room temperature~\cite{Zhang_IEEE2006}.
For both tunnel- and multi-gate RFETs, it has been experimentally demonstrated, that the subthreshold limit value can be decreased down to 30 and 40 mV/dec, respectively~\cite{Gandhi_IEEE2011,Jeon2017,Choi_IEEE2007,Appenzeller93_2004}.


For circuit design, the description of the device behavior based on the nonequilibrium Green’s function (NEGF) method, Wigner transport equation, and Boltzmann  equation formalism is unsuitable in terms of memory and time~\cite{Guo_IEEE2004,Ossaimee_EL2008,Leonard_Nanotech2006,Maneux_SSE2013}.
To reduce the computation time, TCAD simulation tools have been used to analyze the ${I-V}$ characteristics of  RFETs with SB contacts solving the current integral involved in the transport calculations numerically~ \cite{Darbandy2016,Martinie_2012}.
For practical circuit design based on simulations in a SPICE-like environment, compact models are required.
In the framework of the constant effective SB approximation using an energy independent transmission probability, different simple analytical expressions for the drain current have been reported in the literature for RFETs~\cite{Martinie_2012,Jeon2017,Fregonese_IEEE2011,Weber_IEEE2014}.
In these models, the simulated ${I-V}$ characteristics agree with experimental data in a limited bias range~\cite{Maneux_SSE2013}.
The reason is that the analytical expression for the drain current corresponding to the thermionic emission with a shifted Fermi level and including energy-independent transmission can be used at small bias, when the contribution of thermally excited electrons in the total current is large enough~\cite{Bejenari_IEEE2017}.
The analytical current calculations on the basis of drift-diffusion  model do not properly take into account the effect of SB tunneling and BTBT on the electron transport~\cite{Antidormi_IEEE2016,Zhang_2015}.
The empirical continuous compact dc model based on a set of empirical fitting parameters is reliable in the framework of experimental data~\cite{Hasan_2017}, but it cannot be used for predictions.

In this paper, we demonstrate the drain current model, which allows to simplify solving of the current integral. It potentially enables to simulate  ${I-V}$ characteristics of one-dimensional reconfigurable  multil--gate transistors with SB  contacts with  reduced computation time.
We adopt the pseudo-bulk approximation~\cite{Fregonese_IEEE2011} to self--consistently estimate the channel potential variation under applied bias with respect to channel charge.
The  drain--current model captures a number of features such as ballistic transport, transmission through the SB contacts, band-to-band tunneling and ambipolar conduction.
It can be applied to quasi-1D RFETs based on both nanowires and nanotubes at large bias voltages. 

 

\section{Transport Model}
\IEEEpubidadjcol
 \subsection {Energy Band Model} 
We consider $N$ gates with left- and right-end coordinates ${\left[z_{L,n},z_{R,n}\right]}$ ($n=1,2,\dots,N$) placed along the channel. 
The given band model was adopted from the evanescent mode analysis approach~\cite{Oh_IEEE2000,Michetti_IEEE2010,Jimenez_Nanotech2007,Zhang_2015}.
The electrostatic potential, ${\psi(r)}$, inside a transistor contains 
a transverse potential ${\psi_{t}(r)}$, which describes the electrostatics perpendicular to the channel and represents a partial solution of Poisson's equation,
as well as a longitudinal potential ${\psi_{l}(r)}$ called evanescent
mode, responsible for the potential variation along the channel. 
The transverse potential inside the channel is reduced to ${\psi_{t}(r)\approx \psi_{cc}}$, where ${\psi_{cc}}$ is the channel (surface) potential at the current control point~\cite{Lundstrom_IEEE2003,Mothes2015}.
The longitudinal solution ${\psi_{l}(r)}$ is obtained solving the Laplace equation along the transport direction.
Therefore, near the source and drain contacts, the conduction subband edge is  given by exponentially decaying functions. 
Since electrons with high energy mainly tunnel through the Schottky barrier,  the conduction subband edge $E_{\rm{C}}^{\rm{s}}$ ($E_{\rm{C}}^{\rm{d}}$) in the vicinity of the source (drain) contact  can be approximated by a linear decaying function 
\begin{eqnarray}
E_{\rm{C}}^{\rm{s}}(z)=E_{m,0}-q\psi_{cc,1}  + E^s_b\left(1-\frac{z}{\lambda_s}\right), 
\label{eq:SBHight_1} \\
E_{\rm{C}}^{\rm{d}}(z)=E_{m,0}-q\psi_{cc,N}  + E^d_b \left[1+\frac{z-L}{\lambda_{d}}\right],
\label{eq:SBHight_2} 
\end{eqnarray}
where $L$ is the total length of the channel, $\lambda_{s(d)}$ is a characteristic length of the decaying electrostatic potential that can be interpreted as an effective SB width and ${E^{s(d)}_b=\phi_{b}+q\psi_{cc,1(N)}-E_{m,0}-qV_{s(d)}}$ is the bias dependent potential barrier height with respect to the bottom of the $m$th conduction subband ${E_{m,0}-q\psi_{cc,1(N)}}$ at the source and drain contacts, correspondingly.
For cylindrical gate-all-around FETs, the asymptotic value of $\lambda$ is approximately given by ${(2\kappa t_{\rm{ox}}+d_{\rm{ch}})/4.81}$, where ${\kappa=\epsilon_{\rm{ch}}/\epsilon_{\rm{ox}}}$ can be obtained if the oxide thickness, $t_{\rm{ox}}$, is significantly smaller than the channel diameter, ${d_{\rm{ch}}}$~\cite{Oh_IEEE2000}.
For gate-all-around CNTFETs, the CNT diameter, ${d_{\rm{CNT}}}$, is often smaller than the oxide thickness, therefore, the asymptotic value of ${\lambda}$ is slightly modified~\cite{Wong_IEEE2015PI}. In the case of double-gate FETs, the similar approximation of the characteristic length reads ${\lambda\approx (2\kappa t_{\rm{ox}}+t_{\rm{ch}})/\pi}$, where  ${t_{\rm{ch}}}$ is the thickness of the channel~\cite{Oh_IEEE2000}.

Between two adjacent gates with bias voltages ${V_{g,n-1}}$ and ${V_{g,n}}$, the electrostatic potential is supposed to be linearly dependent on space variable in the inner part of the channel. 
Hence, the conduction band edge ${E_{\rm{C,n}}^{\rm{in}}}$ in the nth adjacent interval ${(z_{R,n-1},z_{L,n})}$ is defined as
\begin{multline}
E_{\rm{C,n}}^{\rm{in}}(z)=E_{m,0}-q\psi_{cc,n-1}   \\
+q(\psi_{cc,n}-\psi_{cc,n-1})\left(\frac{z_{R,n-1}-z}{z_{L,n}-z_{R,n-1}}\right), \label{eq:band_inner} 
\end{multline}
where index $n=2,3,\dots,N$.
In the case of mirror-symmetric band structure, the valence subband edge ${E_{\rm{V}}}$ is described as ${E_{\rm{V}}(z)=E_{\rm{C}}(z)-2E_{m,0}}$.

\begin{figure*}[!t]
\centering
\subfloat[]{\includegraphics[width=3.in]{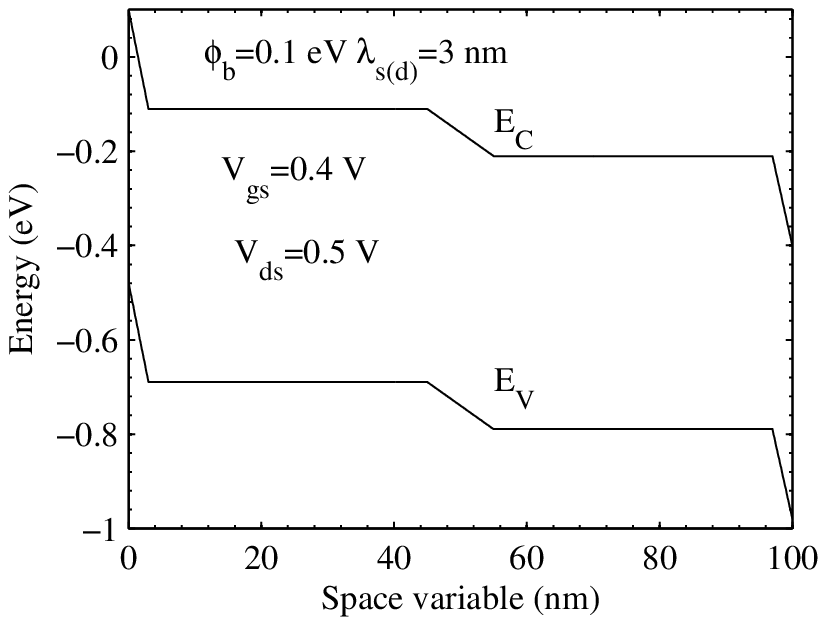}%
\label{fig_first_case}}
\hfil
\subfloat[]{\includegraphics[width=3.in]{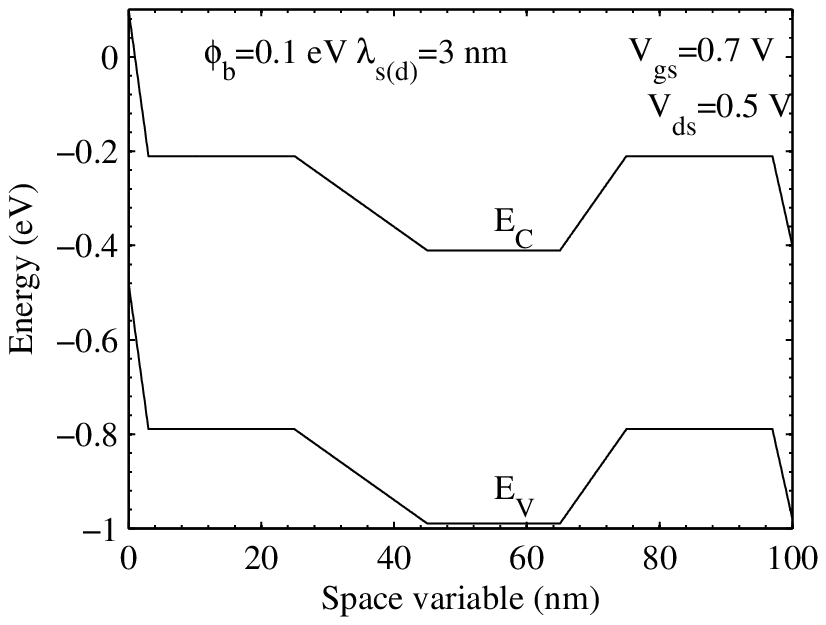}%
\label{fig_second_case}}
\caption{Energy band diagram for (a) n--type double-gate CNTFET with a length of 45 nm of both the 1st and 2nd gates  in the ON-state at the drain--source  and program gate voltages ${V_{ds}=V_{pd}=0.5}$ V, (b) triple-gate CNTFET with a length of 25, 20, and 25 nm of the 1st, 2nd, and 3rd gate, respectively, in the ON-state at the drain--source  and program gate voltages ${V_{ds}=V_{pd}=V_{ps}=0.5}$ V. }
\label{fig1:BandDiagram}
\end{figure*}

Fig.~\ref{fig1:BandDiagram} shows the conduction band profile along the channel.
The gate length ${L_g}$ of the device coincides with the channel length ${L}$. 
The metal-semiconductor barrier height referenced to source Fermi level $E_{Fs}$ is described by a bias independent parameter, ${\phi_{b}}$, which is commonly defined by the difference between the metal work function, $\phi_M$, and semiconductor electron affinity, $\chi_{SC}$, i.e., $\phi_{b}\approx\phi_M-\chi_{SC}$~\cite{Sze_2007,Svensson2011,Tung2014}.
For holes, the similar parameter ${\phi^{h}_{b}}$ is given by ${\phi^{h}_{b}=E_g-\phi_{b}}$, where ${E_g=2E_{m,0}}$ is the band gap.  
The source and drain Fermi levels $E_{Fs}$ and $E_{Fd}$, respectively, are related as $E_{Fd}=E_{Fs}-qV_{ds}$, where $V_{ds}=V_{d}-V_{s}$ is the drain--source voltage.

 
The contribution of electrons injected from the source and drain to the total current depends on both the energy dependent transmission through the channel and electron distribution in the contacts.

\subsection{Piece-Wise Approximation of Fermi-Dirac Distribution Function}

The electron distribution in the source/drain contacts is given by the equilibrium Fermi-Dirac distribution function ${f_{FD}(E-E_F)=1/\left\{\exp\left[\left(E-E_F\right)/{k_BT}\right]+1\right\}}$.
To find an analytical expression for the current, we use a piece-wise approximation for $f_{FD}(E)$ given by~\cite{Bejenari_IEEE2017}
\begin{equation} 
{  f_{\rm{app}}(E)= \left\{
				\begin{array}{l}
					1-\frac{1}{2}\exp\left( \frac{E-E_F}{{c_1 k_BT}}\right),  E \leq E_F  \\
					\frac{1}{2}\exp\left(\frac{E_F-E}{{c_1 k_BT}}\right),~  E_F<E<E_F+c_2 k_BT   \\
				\exp\left(\frac{E_F-E}{{k_BT}}\right), ~   E\geq E_F+c_2 k_BT	
				\end{array}
				\right.\
	\label{eq:Fermi_Dirac}		
}
\end{equation}
where ${c_1=2\ln(2)}$ and ${c_2=2\ln^2(2)/(2\ln(2)-1)\approx 2.49}$. 

The approximation $f_{\rm{app}}(E)$ provides accurate values of the electron distribution function at different temperatures in the whole energy range  with a maximum relative error of about 6-9 percent in the vicinity of Fermi level $E_F$. 

\subsection{Transmission Probability}

In order to estimate the transparency of the source/drain contacts, we use the transmission probability across each SB obtained in the framework of the Wentzel--Kramers--Brillouin (WKB) approximation.
Using the effective mass (parabolic one--band) approach, the probability $T^{s(d)}_{b}(E)$ for electrons to tunnel through a linear decaying potential barrier of the kind ${E^s_b\left(1-z/\lambda\right)}$ or
 ${E^d_b\left[1+(z-L)/\lambda\right]}$ is given by the following expression~\cite{Sze_2007}
\begin{eqnarray}
&T^{s(d)}_{b}(E)  =\exp\left\{ -\alpha \sqrt{|E^{s(d)}_{b}|} \gamma \left(E/|E^{s(d)}_{b}|\right) \right\}, 
\label{eq:Transmission}\\
&\gamma(x)  =\left(1-x\right)^{3/2},
\end{eqnarray}
where ${\alpha=4\lambda_{s(d)}\sqrt{2m^*}/(3\hbar)}$. For CNTs, the electron effective mass is ${m^*=4E_{m,0} \hbar^2/(3a^2V_{\pi}^2)}$ with $a=\SI{2.49}{\angstrom}$ - carbon--carbon atom distance and $V_{\pi}=\SI{3.033}{\electronvolt}$ -- carbon $\pi-\pi$ bond energy in the tight binding model~\cite{Mintmire1998}.

To obtain an analytical expression for the current, we use the following approximation for $\gamma(x)$ in~(\ref{eq:Transmission})
\begin{eqnarray}
&\gamma_{\rm{app}}(x)  =(1-x)(1-px), 
\label{eq:gama_app}\\
&p=\varphi-\sqrt{\varphi^2-\varphi}\approx 0.618,
\end{eqnarray}
where the quantity $\varphi=(1+\sqrt{5})/2$ represents the golden ratio and $x=E/E^{s(d)}_{b}$ is a dimensionless variable.
The absolute error of $\gamma_{\rm{app}}(x)$ is less than 0.016 for all ${x\in[0,1]}$.
Nevertheless, the implementation of $\gamma_{\rm{app}}(x)$ in~(\ref{eq:Transmission}) leads to an increase of relative error of the approximate transmission probability $T^{s(d)}_{\rm{app}}(E)$ with gate voltage due to term ${E^{s(d)}_{b}}$. 
To reduce the relative error, we introduce a correction factor ${\exp\left[\alpha \Delta (E^{s(d)}_{b})^{1/2}\right]}$ with the constant $\Delta < \text{max} \left|\gamma(x)-\gamma_{\rm{app}}(x)\right|$ in the final expression of current.
The approximate transmission probability $T^{s}_{\rm{app}}(E)$ based on~(\ref{eq:Transmission}) and (\ref{eq:gama_app}) is used  in region 2 if there is only one potential barrier. 

If ${E^{s(d)}_{b}>E_{g}}$, electrons can tunnel through the band gap from valence band to conduction band and vice versa.
The probability of such band-to-band (BTB) tunneling for electrons and holes with equal masses is given in the parabolic one-band approximation by
\begin{equation}
T^{s(d)}_{\rm{BTB}} =\exp\left\{ -\alpha E^{3/2}_{g}/ E^{s(d)}_{b}  \right\}.
\label{eq:BTBT}
\end{equation}

In the non-parabolic two-band approximation, the energy dispersion for electrons and holes in CNT is ${E_{m,l}=\pm\sqrt{E_{m,0}^2+(\hbar v_F k_l)^2}}$, where ${v_F\approx10^8}$ cm/s is the Fermi velocity.
The electron effective mass $m^*$ and ${v_F}$ are related by ${m^*=E_{m,0}/v^2_F}$.
In this case, the probability $T^{s(d)}_{b}(E)$ for electrons to tunnel through a linear decaying potential barrier is obtained in the WKB approximation as 
\begin{eqnarray}
&T^{s(d)}_{b}(E)  =\exp\left\{ - \beta \zeta \left(\frac{E+E_{m,0}-|E^{s(d)}_{b}|}{E_{m,0}}\right)/|E^{s(d)}_{b}| \right\}, 
\label{eq:Transmission_2Band}\\
&\zeta(x)  ={\pi}/{2}-x\sqrt{1-x^2} -\arcsin\left(x\right),
\end{eqnarray}
where ${\beta=\lambda_{s(d)} E^2_{m,0}/(\hbar v_F)}$ and $x=E/E^{s(d)}_{b}$ is a dimensionless variable.
In the non-parabolic two-band approximation, the probability of BTB tunneling reads\cite{Jena_APL2008}
\begin{equation}
T^{s(d)}_{\textrm{BTB}} =\exp\left\{ -\frac{3\pi}{16}\alpha E^{3/2}_{g}/ E^{s(d)}_{b}  \right\}.
\label{eq:BTBT_2Band}
\end{equation}
A comparison of (\ref{eq:BTBT}) and (\ref{eq:BTBT_2Band}) shows that the probability of BTB tunneling obtained in the non--parabolic two--band approximation is greater than that obtained in the parabolic one--band approximation. 
In the inner part of the channel,  BTB tunneling ${T^{\rm{in},n}_{\rm{BTB}}}$ of electrons between two adjacent gates is given by (\ref{eq:BTBT}) or (\ref{eq:BTBT_2Band}), where ${E^{s(d)}_{b}}$ is replaced by the difference ${q|\psi_{cc,n}-\psi_{cc,n-1}|}$ and the characteristic length $\lambda_{s(d)}$ is replaced by the distance ${z_{L,n}-z_{R,n-1}}$ between two adjacent gates $n-1$ and $n$ ($n=2,3,\dots,N$). 


If ${q\psi_{cc,1(N)}=E_{m,0}-\phi_{b}+V_{s(d)}}$, there is no SB located at the source (drain), then the transmission probability of electrons or holes to inject from the source (drain) into the channel  is equal to 0 if the electron energy belongs to the band gap (${\phi_b-V_{s(d)} -2 E_{m,0}< E <\phi_b-V_{s(d)}}$) and it is 1 otherwise.  

The probability of electron transmission through the potential barrier increases with energy. 
At a large gate voltage, electrons with high energy or close to the Fermi level tunnel through the thin potential barrier with a rather small reflection probability ${1-T^{s(d)}_{b}(E)}$ and mainly contribute to the current, whereas the contribution of electrons with low energy is not essential due to a small transmission probability ${T^{s(d)}_{b}(E)}$. 
Hence, the multiple reflections between two potential barriers can be neglected.
In this case, the total transmission probability reads
\begin{equation}
T_{tun}(E)=T^{s}(E)T^{in}_{BTB}T^{d}(E).
\label{eq:total_Transmission}
\end{equation}
The approximate total transmission probability ${T^{tun}_{\rm{app}}(E)}$ can be obtained by using (\ref{eq:Transmission})--(\ref{eq:total_Transmission}).

The presented approach is valid if electron-phonon scattering is relatively small, i.e., the channel length $L$ is of the order of an electron mean free path $L_{\rm{mfp}}$, such that ${L/L_{\rm{mfp}}<1/\overline{T}_b}$, where ${\overline{T}_b}$ is an average value of the SB transmission probability characterizing a source/drain contact transparency~\cite{Knoch2008}.
Depending on the applied bias, the mean free path $L_{\rm{mfp}}$ can vary from 60 to 200 nm~\cite{Fuller2014,Franklin_2010,Zhang2008_nl,Purewal2007,Yao2000} at room temperature in CNTFETs. 
Also, the model does not take into account direct source-to-drain tunneling and short--channel effects (e.g., SS degradation and Drain-Induced Barrier Lowering), which are determined purely by electrostatics and essentially affect the current at ${L\approx\lambda}$~\cite{Knoch2008}. 

%

\subsection{Total Current}

To calculate the total electron current, we use the Landauer-Buttiker  approximation for a one-dimensional system~\cite{Datta_1995}
\begin{equation}
I = \frac{4q}{h}  {\int\limits_{-\infty}^\infty T_{tun}(E)\left[ f_{FD}(E-E_{Fs}) - f_{FD}(E-E_{Fd}) \right]dE},
\label{eq:current}
\end{equation}
where the product of the spin and electron subband degeneracies gives a factor of 4 in front of the integral (\ref{eq:current}) for CNTFETs. 

\section{Results}

\begin{figure}[!t]
\centering
\includegraphics[width=3.5in]{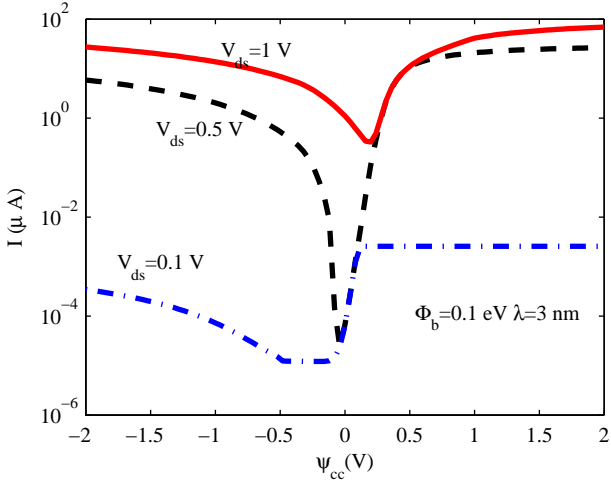}
\caption{Total current $I$  calculated  numerically
as a function of tube potential ${\psi_{cc,1}}$ at the drain--source  and program gate voltages ${V_{ds}=V_{pd}}$ equal to 0.1, 0.5, and 1 V. CNT chirality (19, 0), bandgap ${E_g = 0.579}$ eV, CNT diameter ${d_{\rm{CNT}}=1.48}$ nm, SB hight ${\phi_b}=0.1$ eV, characteristic length ${\lambda_{s(d)}=3}$ nm, equal gate lengths ${L_{g} = 45}$ nm, 
 and temperature ${T = 300}$ K for the two-gate CNTFET. }
\label{fig:I_psi_2G}
\end{figure}

Fig.~\ref{fig:I_psi_2G} shows the total current $I$ calculated  numerically in the framework of two-band approximation (\ref{eq:Transmission_2Band})-(\ref{eq:BTBT_2Band}) and (\ref{eq:current}) as a function of tube potential  ${\psi_{cc,1}}$ at different values of drain--source voltage ${V_{ds}}$ and similar values of the program gate for the two--gate RFET with equal gate lengths of 45 nm.
The program gate voltage  ${V_{pd}}$ is supposed to be equal to the corresponding tube potential ${\psi_{cc,2}}$.

At a larger drain--source voltage (${V_{ds}=2}$ V), the total current strongly depends on the gate voltage in the whole interval of ${\psi_{cc}}$, because the contribution of electrons injected from the drain into the channel is negligibly small  due to the large reflection of such electrons from the potential barrier in the channel.
The On/Off ratio is equal to $2.12\cdot10^2$, ${1.08\cdot10^6}$, and ${2.11\cdot10^2}$ at the drain--source  and program gate voltages ${V_{ds}=V_{pd}}$ equal to 0.1, 0.5, and 1 V, correspondingly.
The subthreshold swing ${SS=\left( {d \log_{10}I}/{d V_{gs}} \right)^{-1}}$  equals 63, 31, and 118 mV/dec, respectively.
Therefore, it can be considerably less than 60 mV/dec at  ${V_{ds}=V_{pd}=0.5}$ V, when the corresponding On/Off ratio is ${1.08\cdot10^6}$.

\begin{figure}[!t]
\centering
\includegraphics[width=3.5in]{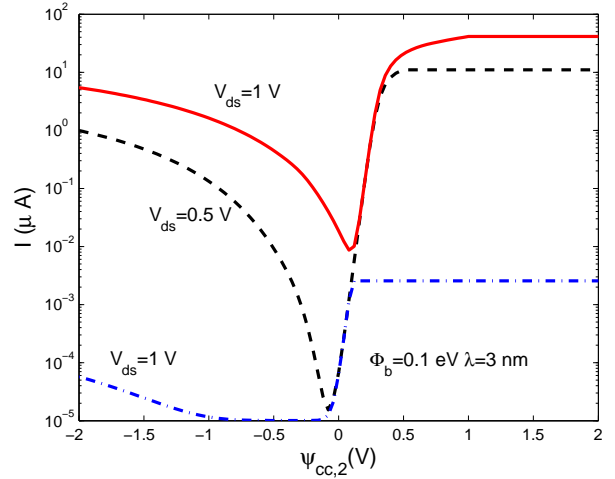}
\caption{Total current $I$  calculated  numerically
as a function of tube potential ${\psi_{cc,2}}$ at the drain--source  and program gate voltages ${V_{ds}=V_{pd}=V_{ps}}$ equal to 0.1, 0.5, and 1 V. CNT chirality (19, 0), bandgap ${E_g = 0.579}$ eV, CNT diameter ${d_{\rm{CNT}}=1.48}$ nm, SB hight ${\phi_b}=0.1$ eV, characteristic length ${\lambda_{s(d)}=3}$ nm,  
 and temperature ${T = 300}$ K for the triple-gate CNTFET with a length of 25, 20, and 25 nm of the 1st, 2nd, and 3rd gate, respectively. }
\label{fig:I_psi_3G}
\end{figure}
Fig.~\ref{fig:I_psi_3G} depicts the total current $I$ calculated  numerically in the framework of two-band approximation (\ref{eq:Transmission_2Band})-(\ref{eq:BTBT_2Band}) and (\ref{eq:current}) as a function of tube potential  ${\psi_{cc,2}}$ at different values of drain--source voltage ${V_{ds}}$ and similar values of the program gates for the triple--gate RFET.
The program gate voltages  ${V_{ps}}$ and ${V_{pd}}$ are supposed to be equal to the values of corresponding tube potentials ${\psi_{cc,1}}$ and ${\psi_{cc,3}}$.
The On/Off ratio is equal to $2.53\cdot10^2$, ${7.17\cdot10^5}$, and ${4.77\cdot10^3}$ at the drain--source  and program gate voltages ${V_{ds}=V_{pd}=V_{ps}}$ equal to 0.1, 0.5, and 1 V, correspondingly.
The subthreshold swing ${SS}$  equals 62, 59, and 63 mV/dec, respectively.
A comparison of transfer characteristics  shown in  Fig.~\ref{fig:I_psi_2G} and Fig.~\ref{fig:I_psi_3G} indicates that the contribution of holes to the current of the n--type triple-gate RFET is diminished by an order of magnitude compared to that of the n--type two-gate RFET, i.e.\, the ambipolarity is greatly reduced. 
As a result, the On/Off ratio for the triple-gate RFET is greater by an order of magnitude in comparison to that for the two-gate CNTFET at large ${V_{ds}}$.
But, the $SS$ is similar to the thermionic limit value of 60 mV/dec, which is about twice greater than ${SS=31}$ mV/dec for the double-gate RFET at ${V_{ds}=V_{pd}=0.5}$ V.

\section{Conclusion}

A simple model for ballistic one-dimensional multi-gate transistors with SB contacts taking into account band-to-band tunneling has been developed.
The model allows to find an analytical solution of the current integral, therefore, it can significantly decrease the evaluation times and eases the implementation of the model in Verilog-A.
We have introduced a piece-wise approximation for Fermi--Dirac distribution function and modified the transmission probability using simple elementary functions, which allow to simplify the current calculations. 
Our model can be used for the analysis of experimental data as well as for performance predictions for different SB heights, characteristic lengths, gate lengths, and either electron effective mass or band gap of channel material for quasi-1D multi--gate RFETs based on both semiconductor nanowires and nanotubes. 
A comparative analysis showed, that the ambipolarity in the triple-gate RFETs is strongly suppressed compared to that in the two-gate RFETs.
In contrast, the subthreshold swing for two-gate RFETs can reach a minimum value of 31 mV/dec, which is about twice less than 60 mV/dec typical for the triple-gate RFETs.  

The author would like to thank Prof. Michael  Schr\"oter and Dr. Martin Claus for valuable discussions.


\ifCLASSOPTIONcaptionsoff
  \newpage
\fi



\bibliographystyle{IEEEtran}

\begin{thebibliography}{}
\providecommand{\url}[1]{#1}
\csname url@samestyle\endcsname
\providecommand{\newblock}{\relax}
\providecommand{\bibinfo}[2]{#2}
\providecommand{\BIBentrySTDinterwordspacing}{\spaceskip=0pt\relax}
\providecommand{\BIBentryALTinterwordstretchfactor}{4}
\providecommand{\BIBentryALTinterwordspacing}{\spaceskip=\fontdimen2\font plus
\BIBentryALTinterwordstretchfactor\fontdimen3\font minus
  \fontdimen4\font\relax}
\providecommand{\BIBforeignlanguage}[2]{{%
\expandafter\ifx\csname l@#1\endcsname\relax
\typeout{** WARNING: IEEEtran.bst: No hyphenation pattern has been}%
\typeout{** loaded for the language `#1'. Using the pattern for}%
\typeout{** the default language instead.}%
\else
\language=\csname l@#1\endcsname
\fi
#2}}
\providecommand{\BIBdecl}{\relax}
\BIBdecl

\end{thebibliography}


\begin{thebibliography}{45}
\providecommand{\url}[1]{#1}
\csname url@samestyle\endcsname
\providecommand{\newblock}{\relax}
\providecommand{\bibinfo}[2]{#2}
\providecommand{\BIBentrySTDinterwordspacing}{\spaceskip=0pt\relax}
\providecommand{\BIBentryALTinterwordstretchfactor}{4}
\providecommand{\BIBentryALTinterwordspacing}{\spaceskip=\fontdimen2\font plus
\BIBentryALTinterwordstretchfactor\fontdimen3\font minus
  \fontdimen4\font\relax}
\providecommand{\BIBforeignlanguage}[2]{{%
\expandafter\ifx\csname l@#1\endcsname\relax
\typeout{** WARNING: IEEEtran.bst: No hyphenation pattern has been}%
\typeout{** loaded for the language `#1'. Using the pattern for}%
\typeout{** the default language instead.}%
\else
\language=\csname l@#1\endcsname
\fi
#2}}
\providecommand{\BIBdecl}{\relax}
\BIBdecl

\bibitem{IRDS}
\BIBentryALTinterwordspacing
\text{International Roadmap for Devices and Systems}. \text{MORE MOORE}.
  \text{White Paper.} 2016 \text{Edition}. [Online]. Available:
  \url{http://irds.ieee.org/}
\BIBentrySTDinterwordspacing

\bibitem{Guo_IEEE2004}
J.~Guo, S.~Datta, and M.~Lundstrom, ``A numerical study of scaling issues for
  \text{Schottky-Barrier} carbon nanotube transistors,'' \emph{IEEE Trans.
  Electron Devices}, vol.~51, no.~2, pp. 172--177, Feb 2004.

\bibitem{Franklin_2012}
A.~D. Franklin, M.~Luisier, S.-J. Han, G.~Tulevski, C.~M. Breslin, L.~Gignac,
  M.~S. Lundstrom, and W.~Haensch, ``Sub-10 nm carbon nanotube transistor,''
  \emph{Nano Lett.}, vol.~12, no.~2, pp. 758--762, Jan 2012.

\bibitem{Peng2017}
C.~Qiu, Z.~Zhang, M.~Xiao, Y.~Yang, D.~Zhong, and L.-M. Peng, ``Scaling carbon
  nanotube complementary transistors to 5-nm gate lengths,'' \emph{Science},
  vol. 355, no. 6322, pp. 271--276, Jan 2017.

\bibitem{Larson_IEEE2006}
J.~M. Larson and J.~P. Snyder, ``Overview and status of metal \text{S/D
  Schottky-Barrier} \text{MOSFET} technology,'' \emph{IEEE Trans. Electron
  Devices}, vol.~53, no.~5, pp. 1048--1058, May 2006.

\bibitem{Leonard2011}
F.~Leonard and A.~A. Talin, ``Electrical contacts to one- and two-dimensional
  nanomaterials,'' \emph{Nature Nanotech.}, vol.~6, no.~12, pp. 773--783, Dec
  2011.

\bibitem{Chen_IEEE2006}
C.~Chen, D.~Xu, E.~Kong, and Y.~Zhang, ``Multichannel carbon-nanotube
  \text{FETs} and complementary logic gates with nanowelded contacts,''
  \emph{IEEE Electron Device Lett.}, vol.~27, no.~10, pp. 852--855, Oct 2006.

\bibitem{Heinze2003}
S.~Heinze, M.~Radosavljevi, J.~Tersoff, and P.~Avouris, ``Unexpected scaling of
  the performance of carbon nanotube \text{Schottky-barrier} transistors,''
  \emph{Phys. Rev. B}, vol.~68, no.~23, p. 235418, Dec 2003.

\bibitem{Chen2005}
Z.~H. Chen, J.~Appenzeller, J.Knoch, Y.~M. Lin, and P.~Avouris, ``The role of
  metal--nanotube contact in the performance of carbon nanotube field-effect
  transistors,'' \emph{Nano Lett.}, vol.~5, no.~7, pp. 1497--1502, Jun 2005.

\bibitem{Weber2017}
W.~M. Weber and T.~Mikolajick, ``Silicon and germanium nanowire electronics:
  physics of conventional and unconventional transistors,'' \emph{Rep. Prog.
  Phys.}, vol.~80, no.~6, p. 066502, Apr 2017.

\bibitem{Mikolajick2017}
T.~Mikolajick, A.~Heinzig, J.~Trommer, T.~Baldauf, and W.~M. Weber, ``The
  {RFET}--a reconfigurable nanowire transistor and its application to novel
  electronic circuits and systems,'' \emph{Semicond. Sci. Technol.}, vol.~32,
  no.~4, p. 043001, Mar 2017.

\bibitem{Zhang_IEEE2006}
Q.~Zhang, W.~Zhao, and A.~Seabaugh, ``Low-subthreshold-swing tunnel
  transistors,'' \emph{IEEE Electron Device Lett.}, vol.~27, no.~4, pp.
  297--300, Apr 2006.

\bibitem{Gandhi_IEEE2011}
R.~Gandhi, Z.~Chen, N.~Singh, K.~Banerjee, and S.~Lee, ``Vertical si-nanowire
  n-type tunneling fets with low subthreshold swing (${<}$ 50 mv/decade) at
  room temperature,'' \emph{IEEE Electron Device Lett.}, vol.~32, no.~4, pp.
  437--439, Apr 2011.

\bibitem{Jeon2017}
D.-Y. Jeon, J.~Zhang, J.~Trommer, S.~J. Park, P.-E. Gaillardon, G.~D. Micheli,
  T.~Mikolajick, and W.~M. Weber, ``Operation regimes and electrical transport
  of steep slope {Schottky} {Si-FinFETs},'' \emph{J. Appl. Phys.}, vol. 121,
  no.~6, p. 064504, Feb 2017.

\bibitem{Choi_IEEE2007}
W.~Y. Choi, B.-G. Park, J.~D. Lee, and T.-J.~K. Liu, ``Tunneling field-effect
  transistors ({TFETs}) with subthreshold swing ({SS}) less than 60 {mV}/dec,''
  \emph{IEEE Electron Device Lett.}, vol.~28, no.~8, pp. 743--745, Aug 2007.

\bibitem{Appenzeller93_2004}
J.~Appenzeller, Y.-M. Lin, J.~Knoch, and P.~Avouris, ``Band-to-band tunneling
  in carbon nanotube field-effect transistors,'' \emph{Phys. Rev. Lett.},
  vol.~93, no.~19, p. 196805, Nov 2004.

\bibitem{Ossaimee_EL2008}
M.~Ossaimee, S.~Gamal, K.~Kirah, and O.~Omar, ``Ballistic transport in
  \text{Schottky} barrier carbon nanotube \text{FETs},'' \emph{Electron.
  Lett.}, vol.~44, no.~5, pp. 336--337, Feb 2008.

\bibitem{Leonard_Nanotech2006}
F.~Leonard and D.~A. Stewart, ``Properties of short channel ballistic carbon
  nanotube transistors with ohmic contacts,'' \emph{Nanotechnology}, vol.~17,
  no.~18, pp. 4699--4705, Aug 2006.

\bibitem{Maneux_SSE2013}
C.~Maneux, S.~Fregonese, T.~Zimmer, S.~Retailleau, H.~N. Nguyen, D.~Querlioz,
  A.~Bournel, P.~Dollfus, F.~Triozon, Y.~M. Niquet, and S.~Roche, ``Multiscale
  simulation of carbon nanotube transistors,'' \emph{Solid-State Electron.},
  vol.~89, pp. 26--67, Nov 2013.

\bibitem{Darbandy2016}
G.~Darbandy, M.~Claus, and M.~Schroeter, ``High-performance reconfigurable {Si}
  nanowire field-effect transistor based on simplified device design,''
  \emph{IEEE Trans. Nanotechnol.}, vol.~15, no.~2, pp. 289--294, Mar 2016.

\bibitem{Martinie_2012}
S.~Martinie, J.~Lacord, O.~Rozeau, C.~Navarro, S.~Barraud, and J.-C. Barbe,
  ``Reconfigurable {FET} {SPICE} model for design evaluation,'' in \emph{Proc.
  Int. Conf. Simulation of Semiconductor Processes and Devices (SISPAD),
  Nuremberg,Germany}, Sep 2016, pp. 165--168.

\bibitem{Fregonese_IEEE2011}
S.~Fregonese, C.~Maneux, and T.~Zimmer, ``A compact model for dual-gate
  one-dimensional {FET}: Application to carbon-nanotube {FETs},'' \emph{IEEE
  Trans. Electron Devices}, vol.~58, no.~1, pp. 206--215, Jan 2011.

\bibitem{Weber_IEEE2014}
W.~M. Weber, A.~Heinzig, J.~Trommer, M.~Grube, F.~Kreupl, and T.~Mikolajick,
  ``Reconfigurable nanowire electronics-enabling a single {CMOS} circuit
  technology,'' \emph{IEEE Trans. Nanotechnol.}, vol.~13, no.~6, pp.
  1020--1028, Nov 2014.

\bibitem{Bejenari_IEEE2017}
I.~Bejenari, M.~Schroter, and M.~Claus, ``Analytical drain current model of
  {1-D} ballistic {Schottky}-barrier transistors,'' \emph{IEEE Trans. Electron
  Devices}, vol.~64, no.~9, pp. 3904--3911, Sep 2017.

\bibitem{Antidormi_IEEE2016}
A.~Antidormi, S.~Frache, M.~Graziano, P.-E. Gaillardon, G.~Piccinini, and G.~D.
  Micheli, ``Computationally efficient multiple-independent-gate device
  model,'' \emph{IEEE Trans. Nanotechnol.}, vol.~15, no.~1, pp. 2--14, Jan
  2016.

\bibitem{Zhang_2015}
J.~Zhang, P.-E. Gaillardon, and G.~D. Micheli, ``A surface potential and
  current model for polarity-controllable silicon nanowire {FETs},'' in
  \emph{Proc. of the 45th European Solid-State Device Research Conference
  (ESSDERC), Graz, Austria}, Sep 2015, pp. 48--51.

\bibitem{Hasan_2017}
M.~Hasan, P.-E. Gatllardon, and B.~Sensale-Rodriguez, ``A continuous compact dc
  model for dual-independent-gate {FinFETs},'' \emph{IEEE J. Electron Devices
  Soc.}, vol.~5, no.~1, pp. 23--31, Jan 2017.

\bibitem{Oh_IEEE2000}
S.-H. Oh, D.~Monroe, and J.~M. Hergenrother, ``Analytic description of
  short-channel effects in fully-depleted double-gate and cylindrical,
  surrounding-gate mosfets,'' \emph{IEEE Electron Device Lett.}, vol.~21,
  no.~9, pp. 445--447, Sep 2000.

\bibitem{Michetti_IEEE2010}
P.~Michetti and G.~Iannaccone, ``Analytical model of one-dimensional
  carbon-based \text{Schottky}-barrier transistors,'' \emph{IEEE Trans.
  Electron Devices}, vol.~57, no.~7, pp. 1616--1625, Jul 2010.

\bibitem{Jimenez_Nanotech2007}
D.~Jimenez, X.~Cartoixa, E.Miranda, J.~Sune, F.~A. Chaves, and S.~Roche, ``A
  simple drain current model for \text{Schottky}-barrier carbon nanotube field
  effect transistors,'' \emph{Nanotechnology}, vol.~18, no.~2, p. 025201, Jan
  2007.

\bibitem{Lundstrom_IEEE2003}
A.~Rahman, J.~Guo, S.~Datta, and M.~S. Lundstrom, ``Theory of ballistic
  nanotransistors,'' \emph{IEEE Trans. Electron Devices}, vol.~50, no.~9, pp.
  1853--1864, Sep 2003.

\bibitem{Mothes2015}
S.~Mothes, M.~Claus, and M.~Schroeter, ``Toward linearity in \text{Schottky}
  barrier \text{CNTFETs},'' \emph{IEEE Trans. Nanotechnol.}, vol.~14, no.~2,
  pp. 372--378, Mar 2015.

\bibitem{Wong_IEEE2015PI}
C.-S. Lee, E.~Pop, A.~D. Franklin, W.~Haensch, and H.-S.~P. Wong, ``A compact
  virtual-source model for carbon nanotube \text{FETs} in the sub-10-nm regime
  -- \text{Part I}: Intrinsic elements,'' \emph{IEEE Trans. Electron Devices},
  vol.~62, no.~9, pp. 3061--3069, Sep 2015.

\bibitem{Sze_2007}
S.~M. Sze and K.~K. Ng, \emph{Physics of Semiconductor Devices}.\hskip 1em plus
  0.5em minus 0.4em\relax John Wiley \& Sons, Inc., 2007.

\bibitem{Svensson2011}
J.~Svensson and E.~E.~B. Campbell, ``Schottky barriers in carbon nanotube-metal
  contacts,'' \emph{J. Appl. Phys.}, vol. 110, no.~11, p. 111101, Nov 2011.

\bibitem{Tung2014}
R.~Tung, ``The physics and chemistry of the \text{Schottky} barrier height,''
  \emph{Appl. Phys. Rev.}, vol.~1, no.~1, p. 011304, Jan 2014.

\bibitem{Mintmire1998}
J.~W. Mintmire and C.~T. White, ``Universal density of states for carbon
  nanotubes,'' \emph{Phys. Rev. Lett.}, vol.~81, no.~12, pp. 2506--2509, Sep
  1998.

\bibitem{Jena_APL2008}
D.~Jena, T.~Fang, Q.~Zhang, and H.~Xing, ``Zener tunneling in semiconducting
  nanotube and graphene nanoribbon p-n junctions,'' \emph{Appl. Phys. Lett.},
  vol.~93, no.~11, p. 112106, Nov 2008.

\bibitem{Knoch2008}
J.~Knoch and J.~Appenzeller, ``Tunneling phenomena in carbon nanotube
  field-effect transistors,'' \emph{Phys. Stat. Sol. (A)}, vol. 205, no.~4, pp.
  679--694, Mar 2008.

\bibitem{Fuller2014}
E.~J. Fuller, D.~Pan, B.~L. Corso, O.~T. Gul, and P.~G. Collins, ``Mean free
  paths in single-walled carbon nanotubes measured by \text{Kelvin} probe force
  microscopy,'' \emph{Phys. Rev. B}, vol.~89, no.~24, p. 245450, Dec 2014.

\bibitem{Franklin_2010}
A.~D. Franklin and Z.~Chen, ``Length scaling of carbon nanotube transistors,''
  \emph{Nature Nanotechnol.}, vol.~5, no.~12, pp. 858--862, Dec 2010.

\bibitem{Zhang2008_nl}
Z.~Zhang, S.~Wang, L.~Ding, X.~Liang, T.~Pei, J.~Shen, H.~Xu, Q.~Chen, R.~Cui,
  Y.~Li, and L.-M. Peng, ``Self-aligned ballistic n-type single-walled carbon
  nanotube field-effect transistors with adjustable threshold voltage,''
  \emph{Nano Lett.}, vol.~8, no.~11, pp. 3696--3701, Nov 2008.

\bibitem{Purewal2007}
M.~Purewal, B.~Hong, A.~Ravi, B.~Chandra, J.~Hone, and P.~Kim, ``Scaling of
  resistance and electron mean free path of single-walled carbon nanotubes,''
  \emph{Phys. Rev. Lett.}, vol.~98, no.~12, p. 186808, May 2007.

\bibitem{Yao2000}
Z.~Yao, C.~L. Kane, and C.~Dekker, ``High-field electrical transport in
  single-wall carbon nanotubes,'' \emph{Phys. Rev. Lett.}, vol.~84, no.~13, pp.
  2941--2944, Mar 2000.

\bibitem{Datta_1995}
S.~Datta, \emph{Electronic Transport in Mesoscopic Systems}.\hskip 1em plus
  0.5em minus 0.4em\relax Cambridge University Press, 1995.


\end{thebibliography}
%



\providecommand{\noopsort}[1]{}\providecommand{\singleletter}[1]{#1}%

%








\end{document}